\documentclass{PoS}
\usepackage[percent]{overpic}
\title{Air shower reconstruction using HAWC and the Outrigger array}

\ShortTitle{Air shower reconstruction using HAWC and the Outrigger array}

\author{\speaker{Vikas Joshi}$^{ab}$, Harm Schoorlemmer$^a$ for the HAWC Collaboration$^c$\footnote{for collaboration list, see PoS(ICRC2019) 1177.}
\\$^a$Max Planck Institut f\"{u}r Kernphysik, Heidelberg, Germany \\$^b$now at Friedrich-Alexander-Universit\"{a}t Erlangen-N\"{u}rnberg, Erlangen Centre for Astroparticle Physics, Erlangen, Germany \\$^c$For a complete author list, see \textcolor{blue}{https://www.hawc-observatory.org/collaboration/icrc2019.php} \\E-mail: \email{vikas.joshi@fau.de}
\\  \hspace{11mm}      \email{harmscho@mpi-hd.mpg.de}}

\abstract{The High Altitude Water Cherenkov (HAWC) gamma-ray observatory detects cosmic- and gamma-ray initiated air showers in the TeV energy range using 300 water Cherenkov detectors (WCDs). To improve its sensitivity at the highest energies, HAWC has been upgraded with a sparse array of 345 small WCDs (outrigger array) around the HAWC main array. The outrigger array increases the instrumented area of HAWC by a factor of 4 and has started taking data since August 2018. A new gamma-ray reconstruction method has been developed to improve the reconstruction of the air showers which combines the data of mixed type particle detector arrays. In this contribution, we will show the first results of the combined air shower reconstruction of HAWC and its outrigger array using Monte Carlo simulations and the first combined experimental data set.}

\FullConference{36th International Cosmic Ray Conference -ICRC2019-\\
		July 24th - August 1st, 2019\\
		Madison, WI, U.S.A.}

\begin{document}

\section{Introduction}
\begin{figure}[h!]
\centering
\includegraphics[width=\linewidth]{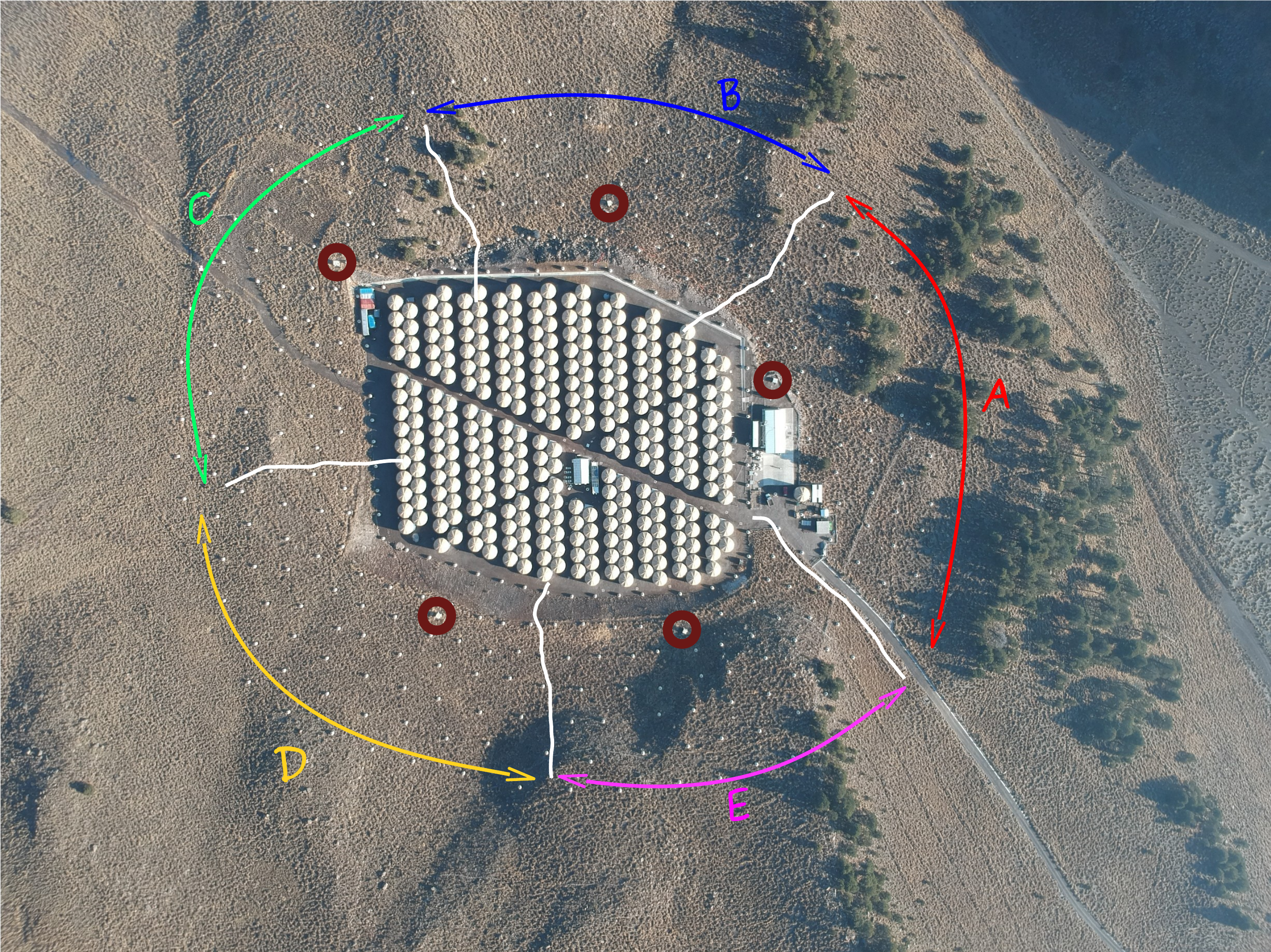}
\caption[Fully deployed outrigger array.]{Fully deployed outrigger array (small tanks) around the main HAWC array. The white lines divide the outrigger array in different sections (A, B, C, D, and E). The dark red circles show the node locations hosting the trigger and readout electronics for their respective outrigger section.}
\label{fig:HAWC+OR_real_image}       
\end{figure}

 The High Altitude Water Cherenkov gamma-ray observatory (HAWC) \cite{hawc_sensitivity} is a second generation ground-based Water Cherenkov Detector (WCD) array. It is located at Sierra Negra in the state of Puebla in central Mexico (18$^\circ$ 59'41" N, 97$^\circ$18'30.6" W) at 4100 m above sea level. The array is composed of 300 WCDs encompassing an area of 22000 m$^2$ (see Figure \ref{fig:HAWC+OR_real_image}, the central array). HAWC is sensitive to gamma- and cosmic-rays in an energy range of a few hundreds of GeV to $\sim$100 TeV energies. The observatory has a large instantaneous field of view of $\sim$2\,sr which covers the declination band of -24$^\circ$ to 64$^\circ$ on a daily basis. 
 It has been fully operational since March 2015 with a duty cycle >95\%.


HAWC is an excellent instrument to study the gamma-ray sky at the highest energies. The study of sources emitting gamma-rays above a few tens TeV energy are of great interest, because these sources might be associated to PeVatrons that accelerate cosmic rays to PeV energies.  Additionally, the study of diffuse emission or extended sources at these energies may shed light to more exotic phenomena.

\section{The Outrigger Array}
\label{outrigger array}
HAWC has been recently upgraded with a sparse array of smaller WCDs around the main HAWC array. This is because, the footprint of the shower on the ground is inherently dependent on the primary particle energy and on the altitude of the detector plane. At HAWC altitude, the footprint of the shower at around $\sim$10 TeV primary particle energy becomes comparable to the total instrumented area. Therefore, most of the showers at these energies are not well-contained within the array. Although the
HAWC main array still has enough information to perform gamma-hadron separation, direction reconstruction, and shower size estimation, the shower reconstruction suffers due to the large uncertainty in the core location. Using the outrigger array, it will be possible to better constrain the core location, so that the shower reconstruction can be improved. It will lead to an increased number of well-reconstructed showers above multi-TeV energies. Hence, it will improve the sensitivity of HAWC at those energies.

HAWC outrigger array consists of 345 cylindrical tanks of diameter 1.55 m and height 1.65 m. Each with one 8" PMT anchored at the bottom of the tank \cite{outriggers_vincent,OR_icrc2017}. The outrigger array is deployed in a concentric circular symmetric way around the main array (see Figure \ref{fig:HAWC+OR_real_image}). The
outrigger array increases the instrumented area of HAWC by a factor of 4-5. The outriggers are mutually separated from each other by 12 to 18 m. The smaller size and larger separation of
the outrigger WCDs are prompted by the fact that there are a lot of particles and consequently bigger signals present near to the core of a big shower. For trigger and readout purpose, the
outrigger array is divided into 5 sections of 70 outriggers each, connected to a node with equal cable lengths. Each node hosts the power supply and the trigger, readout, and calibration system
for the corresponding section. The fully deployed outrigger array has started taking data since August 2018.

\section{Air Shower Reconstruction}
\label{air shower reconstruction}
To combine the reconstruction of the Outrigger array with the one of the HAWC main array and to improve the reconstruction of the HAWC in general, a new Monte Carlo (MC) template-based reconstruction method has been developed with a focus on shower core and energy reconstruction of $\gamma$-ray induced air
showers \cite{template_method}. The algorithm fits an observed lateral amplitude distribution function (LDF) of an extensive air shower against an expected probability distribution function (PDF) using a likelihood approach for a given shower arrival direction. In this case, the LDF is the observed number of photo-electrons (N$_{\rm pe}$) at a given distance from the shower axis (r). A full MC air shower simulation in combination with the HAWC detector simulation (HAWCSim) \cite{crab_paper_hawc, geant4} is used to generate the expected PDFs. The PDFs are binned in energy, X$_{\rm max}$ and zenith angle bins. Each such PDF template is further binned N$_{\rm pe}$ and r. The probability of an observed LDF deduced from the PDF is used to construct the negative log-likelihood, which is then minimised to estimate the best describing fit parameter values, such as energy and the core location. 

\begin{figure}[!h]
\centering
\includegraphics[trim=0.2cm 0.3cm 0.8cm 0.2cm, clip,width=0.97\linewidth]{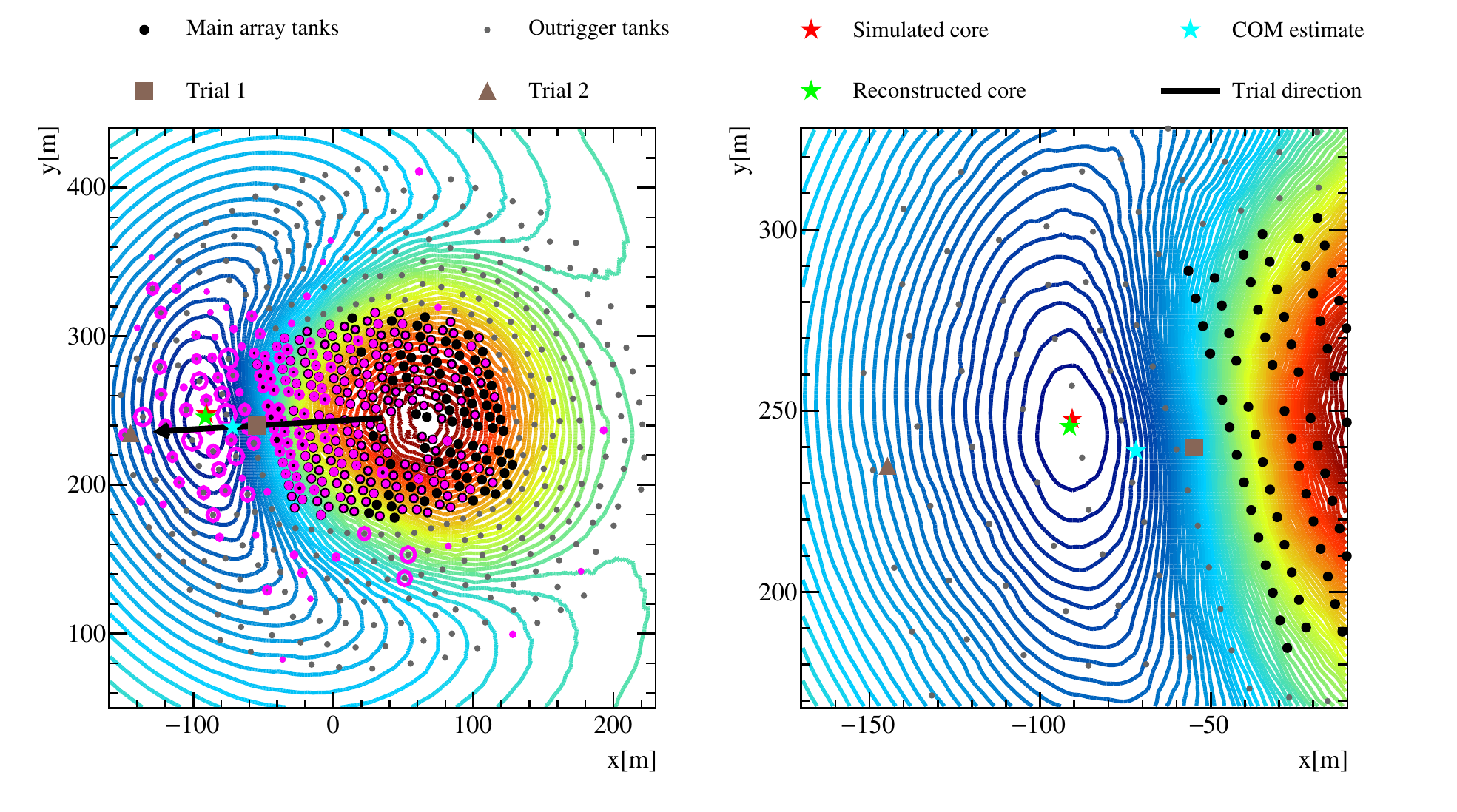}
\caption{The Figure on the right is the zoom in version of the figure on the left around the reconstructed core. Blue to red colour contours show the minimum and the maximum of the likelihood surface respectively. The magenta circles over the tanks show the relative charge observed between the different tanks. The Center-Of-Mass (COM) estimate is calculated using the observed signal amplitudes. The Figure is reproduced from \cite{template_method}.}
\label{fig:likelihood_surface_HAWC+OR}
\end{figure}

\begin{figure}[!h]
\centering
\includegraphics[trim=0.2cm 1.0cm 0.5cm 1.5cm, clip,width=0.495\linewidth]{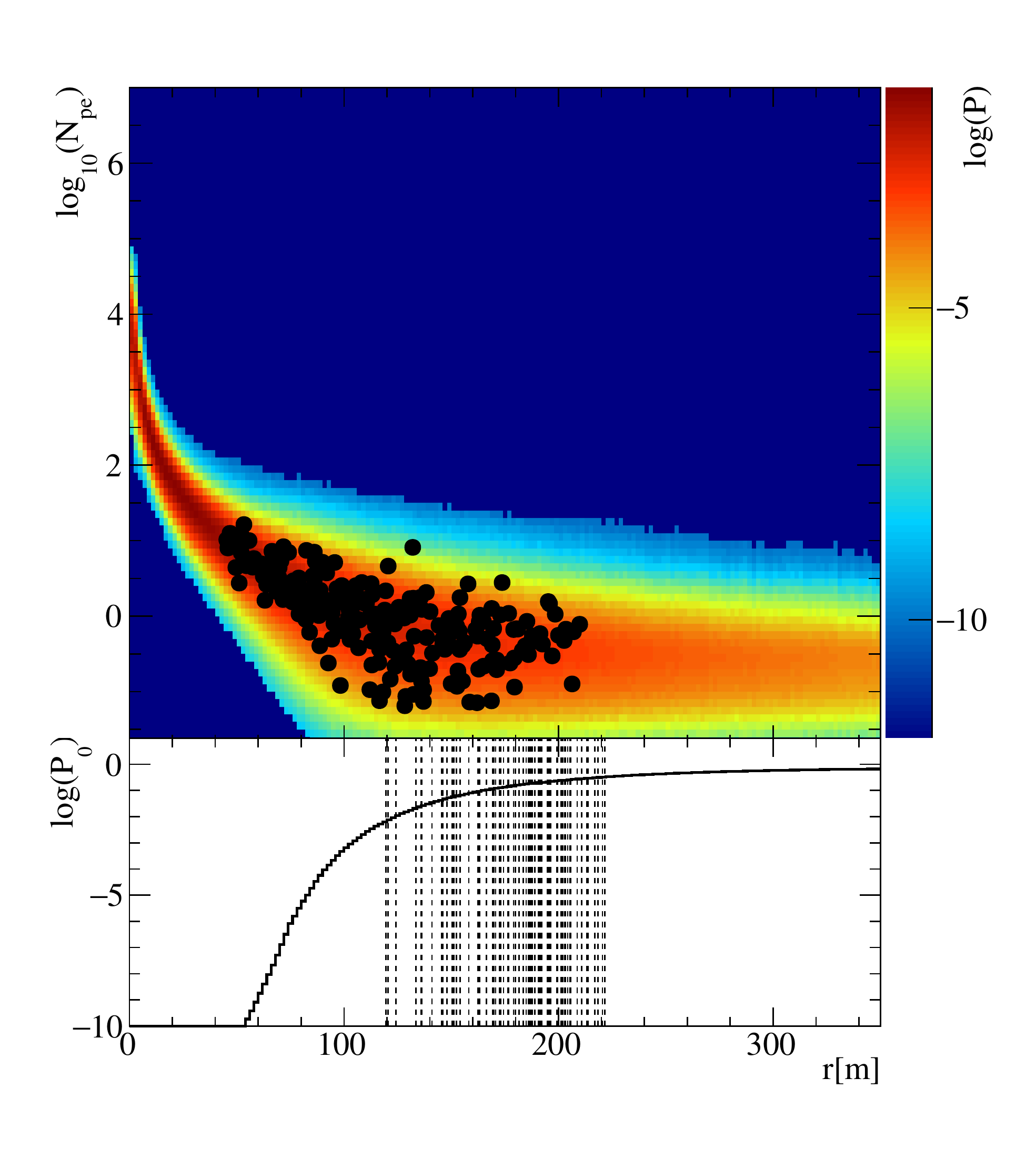}
\includegraphics[trim=0.2cm 1.0cm 0.5cm 1.5cm, clip,width=0.495\linewidth]{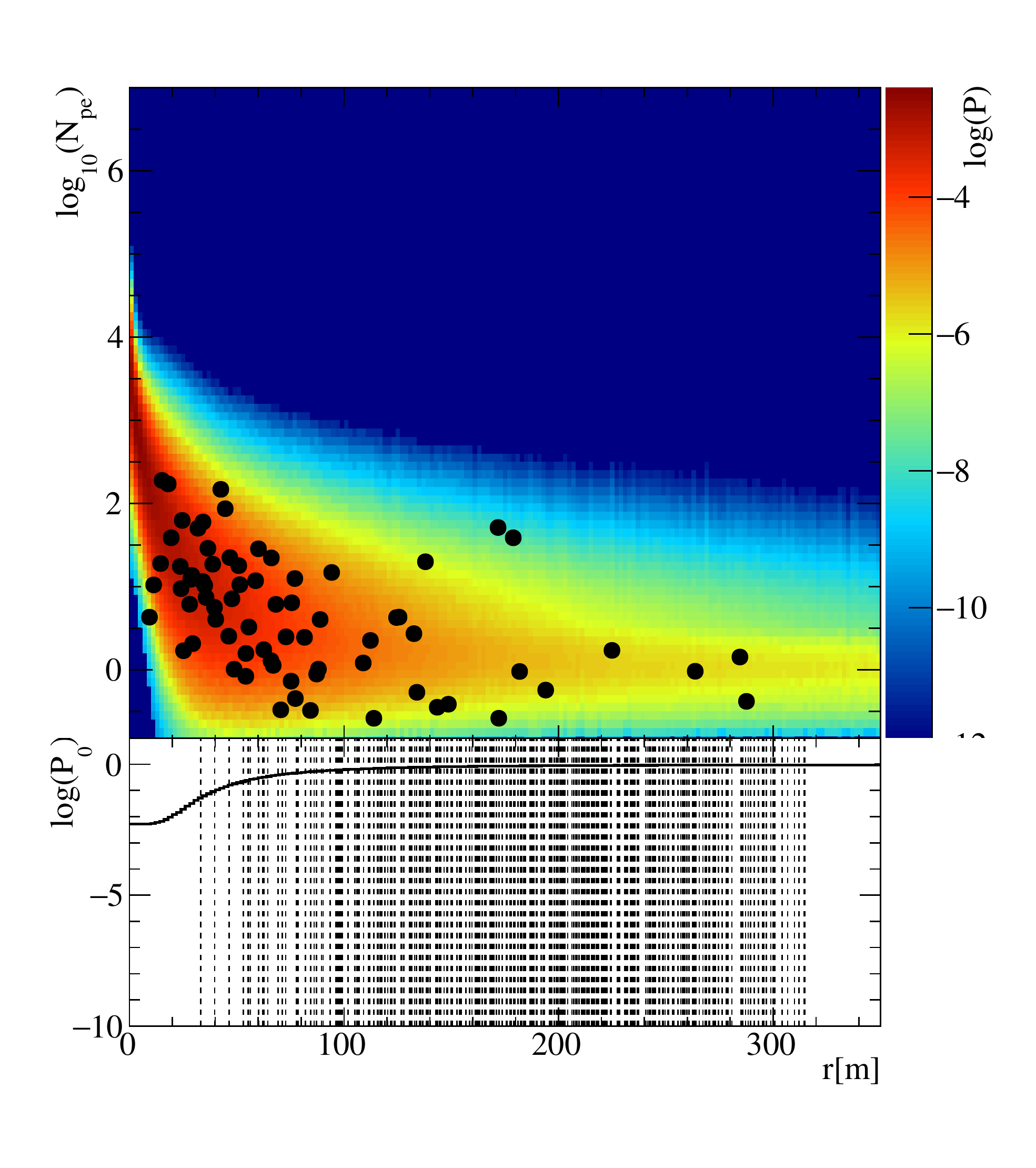}
\caption{The LDF and PDF templates (main array tanks: left, outrigger tanks: right) corresponding to the simulated event shown in the Figure \ref{fig:likelihood_surface_HAWC+OR}, with true energy $\sim$17 TeV and reconstructed energy of $\sim$21 TeV and true X$_{\rm max}$ $\sim$475 g/cm$^{2}$ and reconstructed X$_{\rm max}$  of $\sim$430 g/cm$^{2}$ and zenith angle of 18.17$^{\circ}$. The black dots show the LDF for non-zero N$_{\rm pe}$ and observed zeros are shown as the dashed lines on the histograms below with their corresponding probability (P and P$_0$).  The figure is reproduced from \cite{template_method}.}
\label{fig:LDF_PDF_HAWC+OR}
\end{figure}
The working of the method for the combined reconstruction of the HAWC main array with the Outrigger array is illustrated with one example MC event shown in Figures \ref{fig:likelihood_surface_HAWC+OR}. The signals are detected in both WCDs, the outriggers and the HAWC main array. It shows the likelihood contours and the reconstructed and true core for a simulated event located in the area instrumented by the outrigger array.  The red and green colour stars show the true (simulated) and reconstructed core locations respectively. The reconstructed core location is coincident with  the prominent minimum shown with blue coloured contours of the likelihood surface. Therefore, it can be seen that the method converged to very close to the true core location.

For this event, in Figure \ref{fig:LDF_PDF_HAWC+OR} the best fitting templates corresponding to the main array tanks and the outriggers are shown in the left and right panel respectively. In the case of the smaller outrigger detectors, the fluctuations in the particle density of the shower front become more prominent, which manifest in both larger amplitude
fluctuations and more zero signal detectors. Nevertheless, it is not a problem for the template-based likelihood fit procedure, which combines these significantly different detector responses
naturally to reconstruct the $\gamma$-ray properties, such as in this case, the core location and the energy of the primary particle.

\section{Simulation results}

\subsection{Core Estimation}
\label{core estimation}
\begin{figure}[!h]
\centering
\begin{overpic}[trim=1.0cm 0.5cm 1.0cm 2.0cm, clip,width=0.8\linewidth]{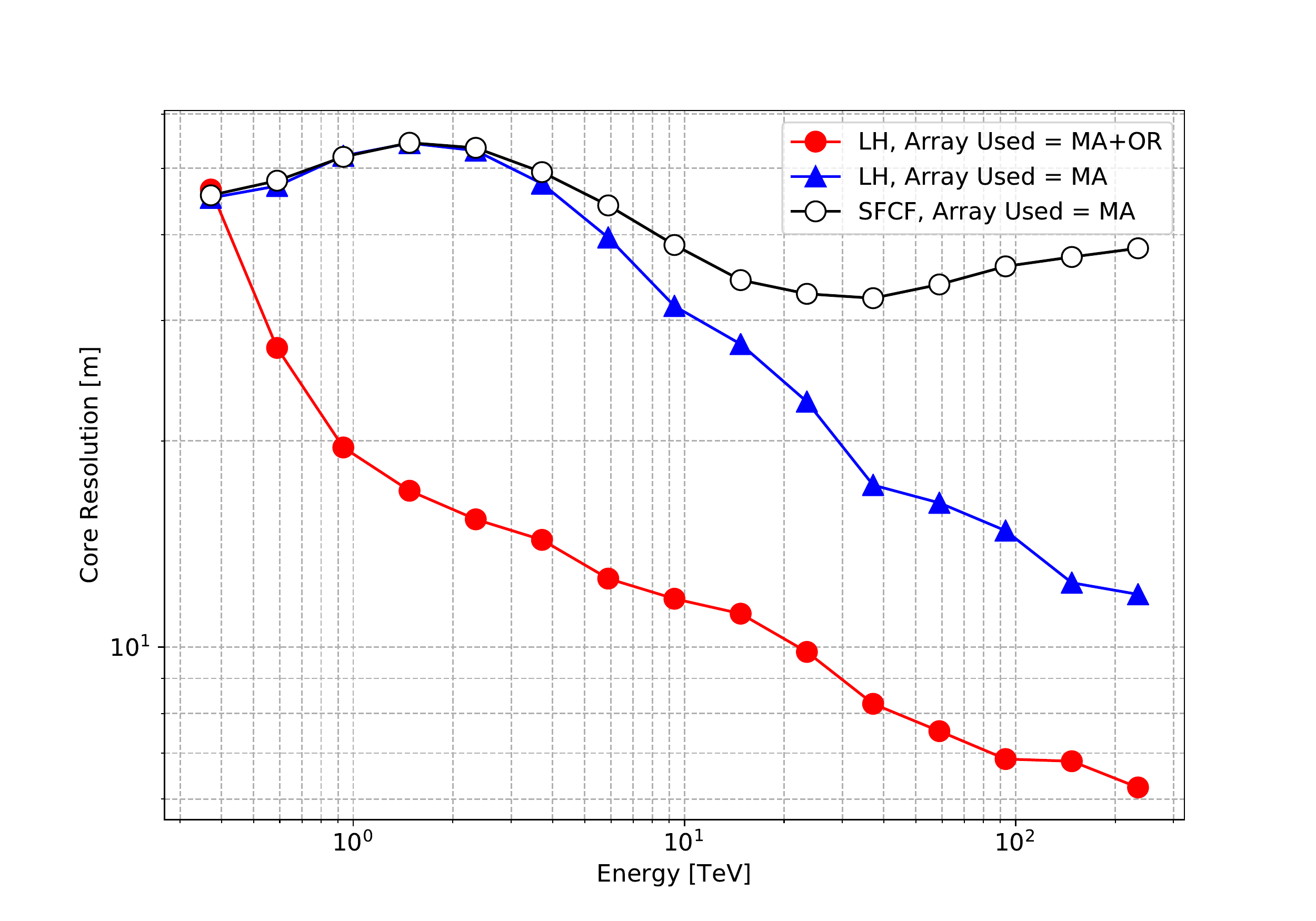}
\put(40,40){\textcolor{black}{\textbf{Preliminary}}}
\end{overpic}
\caption{Core resolution (68\% containment radius) shown as a function of the true energy of the $\gamma$-ray photon. Here SFCF and LH represent the result with HAWC present core estimator and this likelihood fit method respectively. The result is shown for the events falling on the Outrigger array only. Array used means the array used for reconstruction. MA and OR stand for the HAWC main array and outriggers respectively.}
\label{fig:result_core_reso_OR}
\end{figure}

In Figure \ref{fig:result_core_reso_OR}, the expected improvement in the core resolution due to the outrigger array is shown as a function of true energy for the $\gamma$-ray induced air showers.

To obtain the core resolution only the showers falling on the area of the Outrigger array are taken into account. The core
resolution is defined as 68\% containment radius of the distribution of the distance between the reconstructed and true shower core. In order to have enough outrigger detectors triggered,
an additional condition of at least 4\% of the Outrigger array trigger was applied. To evaluate the improvement due to the outrigger array, the comparison is shown between the reconstruction of the events falling on the area of the outrigger array with only HAWC main array (MA) using the SFCF (present core estimator employed in HAWC reconstruction)
and LH (likelihood) method, and with that using the HAWC main array and the Outrigger array (OR) combined for
the LH method. It can be seen that the outrigger array improves the core resolution by $\sim$3 fold in comparison to SFCF above 1 TeV energies. Although it is to be noted that the LH method already performs better than the SFCF method even without using the outrigger array. However, the core resolution improves further for the LH method while using the main array and outrigger array combined. The improvement is about 3 fold around 10 TeV energies and drops down to 2 fold at the highest energies. This better core resolution will therefore result in improvement in the reconstruction of other shower properties such as arrival direction, energy estimation and gamma-hadron separation.  

\subsection{Energy Estimation}
\begin{figure}[!h]
\centering
\begin{overpic}[trim=0.5cm 0.5cm 1.0cm 1.5cm, clip,width=0.7\linewidth]{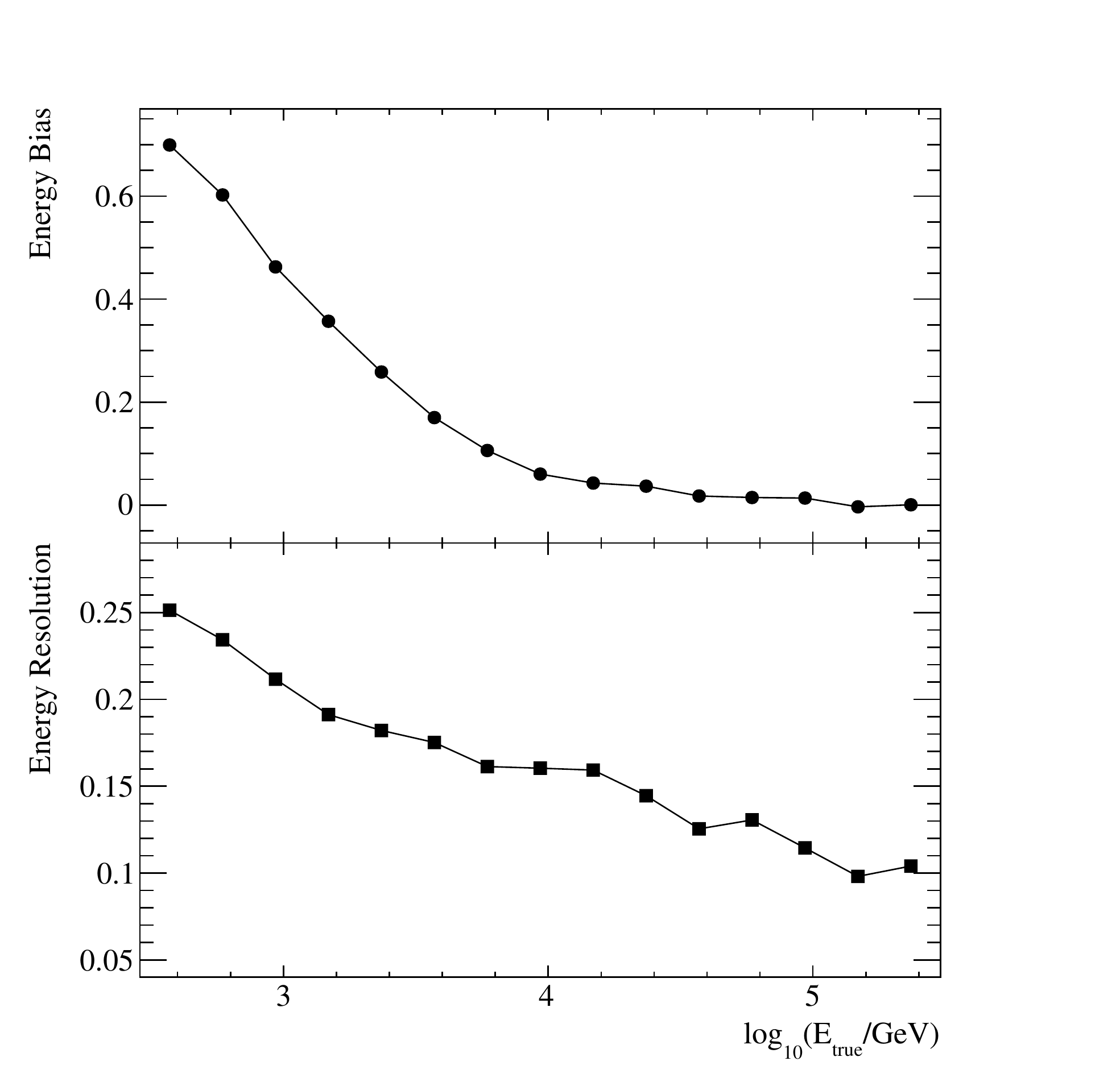}
\put(40,40){\textcolor{black}{\textbf{Preliminary}}}
\end{overpic}
\caption{Fractional energy bias (top) and energy resolution (bottom) as a function of true $\gamma$-ray photon
energy. The results are shown for the combined reconstruction of the main and the outrigger array for the events falling on the Outrigger array.}
\label{fig:result_energy_res}
\end{figure}

The performance of the energy reconstruction can be evaluated by the fractional deviation ($\log_{\rm 10}$(E$_{\rm reco}$) - $\log_{\rm 10}$(E$_{\rm true}$)) of the reconstructed energy with respect to the true energy. The deviation of the mean of the fractional deviation distribution from zero (see Figure \ref{fig:result_energy_res}, top panel)
is the bias in the energy reconstruction. Similarly, the RMS can be understood as the energy resolution (see Figure \ref{fig:result_energy_res}, bottom panel).
The results are shown for the showers falling on the Outrigger array (same events used in the core resolution study shown in section \ref{core estimation}) based on the method described in \cite{template_method} and they are therefore different from the results shown in \cite{energy_estimator_paper}. A comparison of the performance of energy estimation methods used in \cite{template_method} and \cite{energy_estimator_paper} are beyond the scope of this contribution.

The energy bias is large at the low energies but converges to zero at energies >10 TeV, which indicates the stable region of the energy estimation whilst combining the HAWC main array and the outrigger reconstruction. It is to be noted that the expected improvement due to outriggers will be in the energy range above 10 TeV as discussed in section \ref{outrigger array} for the showers falling in the Outrigger array. The bottom panel of Figure \ref{fig:result_energy_res} shows the energy resolution of the likelihood fit method, which starts at $\sim$50\% at 10 TeV energies and improves to $\sim$25\% at the highest energies. 

\section{Experimental Data}
\begin{figure}[!h]
\centering
\includegraphics[trim=0.2cm 0.3cm 0.8cm 0.2cm, clip,width=0.98\linewidth]{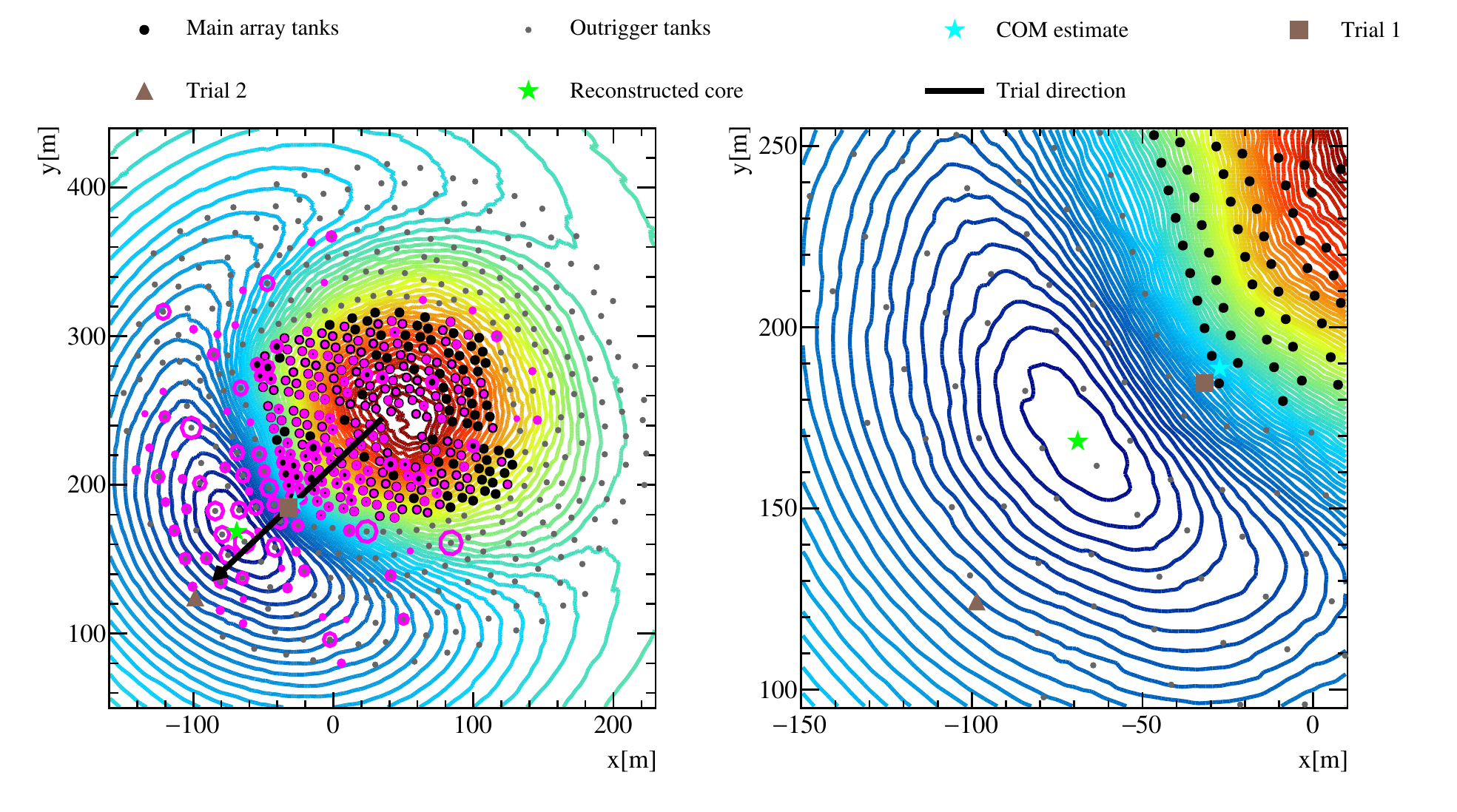}
\caption{Similar to the figure \ref{fig:likelihood_surface_HAWC+OR} but for a $\gamma$-ray like data event coming from the vicinity of the Crab Nebula is shown.}
\label{fig:likelihood_surface_HAWC+OR_data}
\end{figure}

\begin{figure}[!h]
\centering
\includegraphics[trim=0.2cm 1.0cm 0.5cm 1.5cm, clip,width=0.495\linewidth]{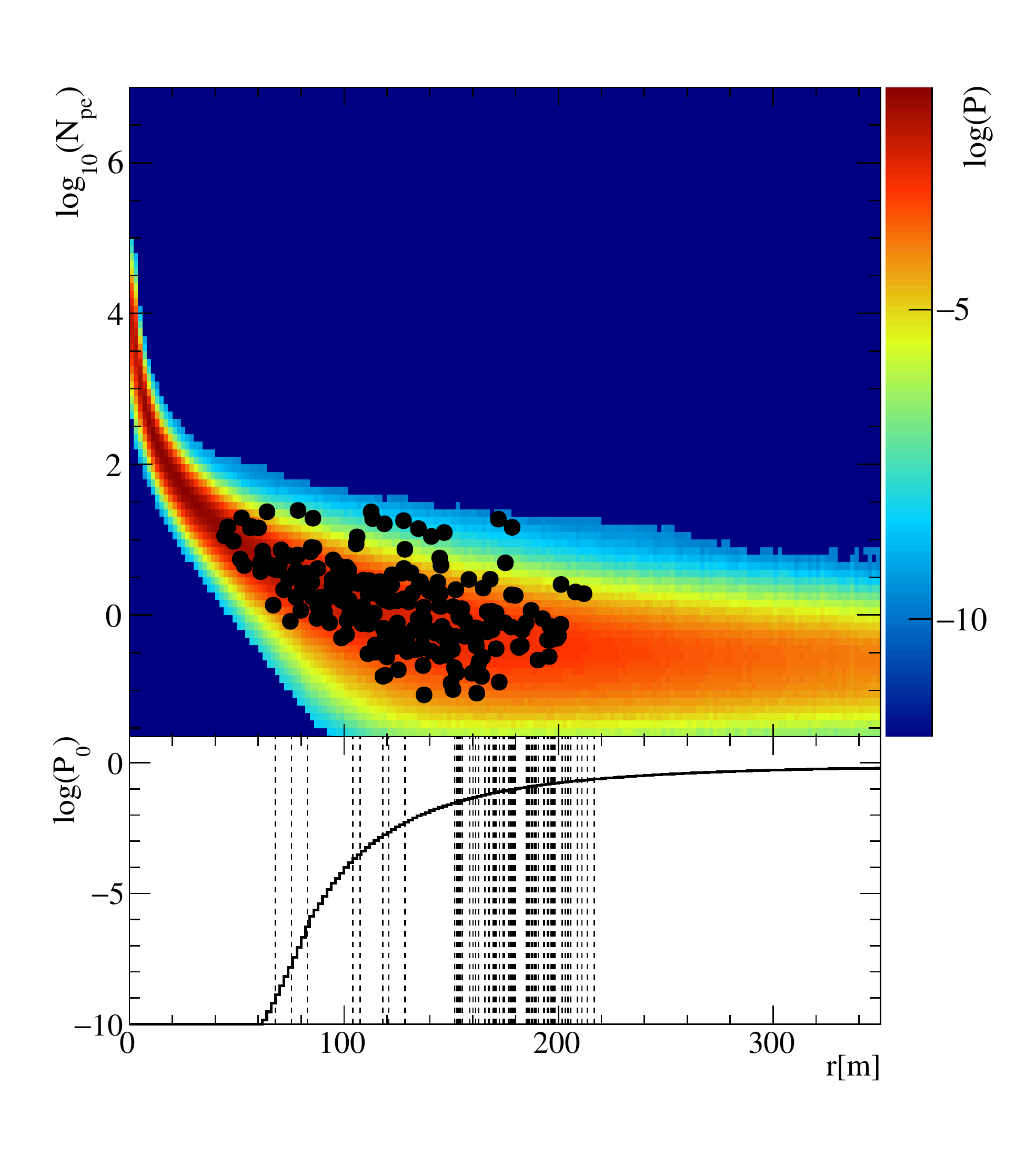}
\includegraphics[trim=0.2cm 1.0cm 0.5cm 1.5cm, clip,width=0.495\linewidth]{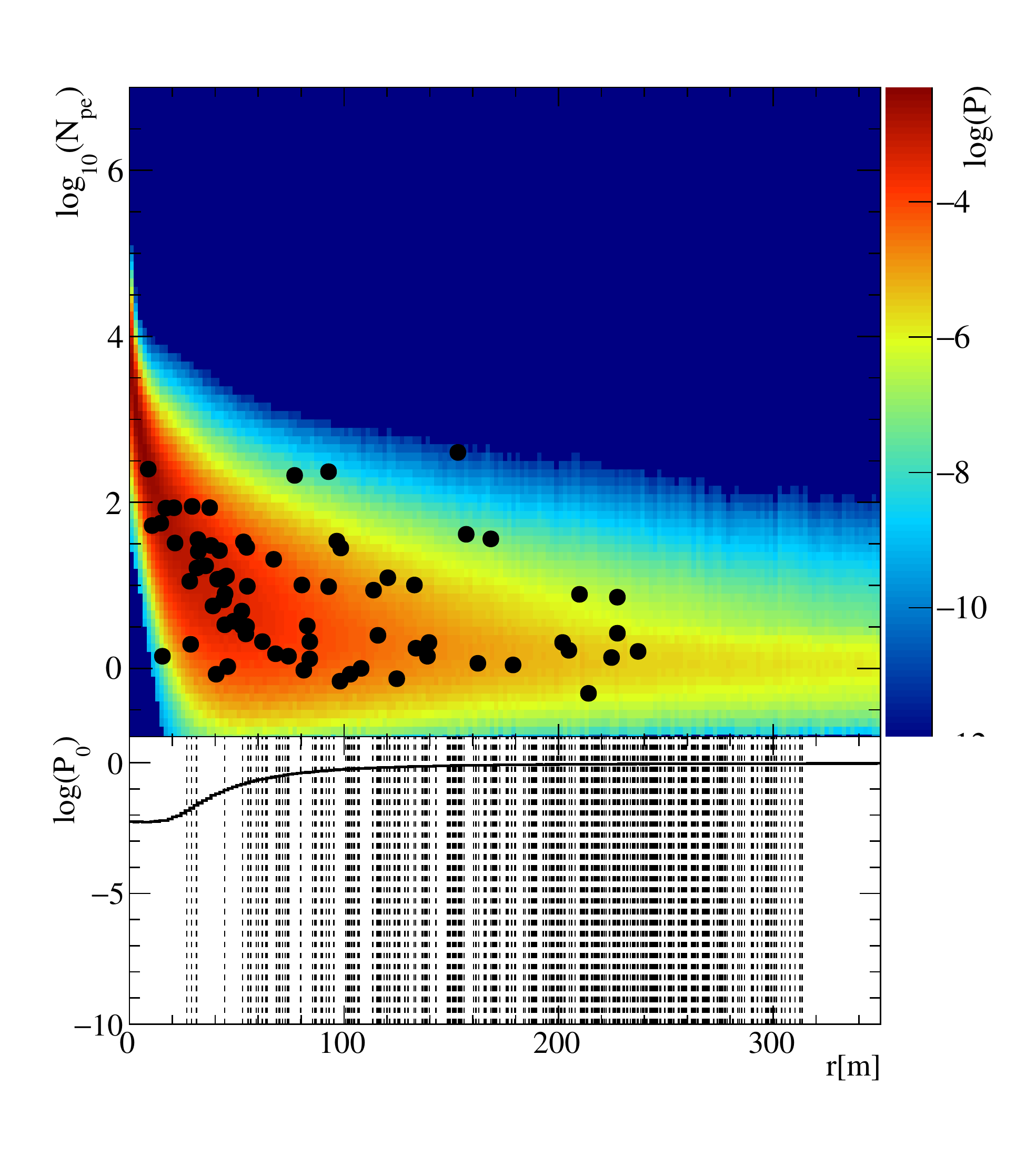}
\caption{Description is similar to Figure \ref{fig:LDF_PDF_HAWC+OR} corresponding to the Crab Nebula data event shown in  figure \ref{fig:likelihood_surface_HAWC+OR_data}, with reconstructed energy $\sim$27 TeV, X$_{\rm max}$ $\sim$425 g/cm$^{2}$ and zenith angle 16.08$^{\circ}$.}
\label{fig:LDF_PDF_HAWC+OR_data}
\end{figure}

To illustrate the proof of the concept, in Figure \ref{fig:likelihood_surface_HAWC+OR_data}, an example candidate $\gamma$-ray event coming from the vicinity of the Crab Nebula and observed using both the HAWC main array and the Outrigger array is shown. In Figure \ref{fig:LDF_PDF_HAWC+OR_data}, the corresponding LDF and PDF are shown. The explanation of the likelihood surface and the LDF and PDF of the given event is the same as discussed in Section \ref{air shower reconstruction}. 
However, it is to be noted that the hit selection and more realistic outrigger simulation is still a work in progress. Therefore, this Crab Nebula event is selected using the existing criteria for the main array only. Nevertheless, it can be seen that the new likelihood method reconstructs the event by fitting it to a PDF that closely represents the LDF of the observed data event.

\section{Outlook}
In this contribution, the combined reconstruction of the HAWC main and the outrigger array is shown by utilising the newly developed MC template-based likelihood fit method. The simulation result on the core estimation improvement is very promising together with another tool to perform the energy estimation with the outrigger array. The first look to the application of the new reconstruction method on the experimental data-set looks encouraging. Currently efforts are being made within the HAWC collaboration to fully integrate the outrigger array for the science operation together with the main array. Those include the evaluation of the improvement in the sensitivity due to outriggers and to understand and utilise the combined experimental data-set of the HAWC main and the outrigger arrays.  

\section*{Acknowledgements}

We acknowledge the support from: the US National Science Foundation (NSF); the US Department of Energy Office of High-Energy Physics; the Laboratory Directed Research and Development (LDRD) program of Los Alamos National Laboratory; Consejo Nacional de Ciencia y Tecnolog\'{i}a (CONACyT), M\'{e}xico (grants 271051, 232656, 260378, 179588, 254964, 258865, 243290, 132197) (C\'{a}tedras 873, 1563, 341), Laboratorio Nacional HAWC de rayos gamma; L'OREAL Fellowship for Women in Science 2014; Red HAWC, M\'{e}xico; DGAPA-UNAM (grants AG100317, IN111315, IN111716-3, IA102715, IN111419, IA102019, IN112218), VIEP-BUAP; PIFI 2012, 2013, PROFOCIE 2014, 2015; the University of Wisconsin Alumni Research Foundation; the Institute of Geophysics, Planetary Physics, and Signatures at Los Alamos National Laboratory; Polish Science Centre grant DEC-2014/13/B/ST9/945, DEC-2017/27/B/ST9/02272; Coordinaci\'{o}n de la Investigaci\'{o}n Cient\'{i}fica de la Universidad Michoacana; Royal Society - Newton Advanced Fellowship 180385. Thanks to Scott Delay, Luciano D\'{i}az and Eduardo Murrieta for technical support.

\end{document}